\documentclass[useAMS,usenatbib]{mn2e}
\usepackage[dvips]{graphicx}
\usepackage[english]{babel}
\usepackage{amsmath}
\usepackage{amssymb}
\usepackage{textcomp}
\usepackage{natbib}

\title[Cold flows as H\,{\sc i} absorption systems]{Cold accretion flows and the nature of high column density H\,{\sc i} absorption at redshift 3}
\author[F. van de Voort et al.]{Freeke van de Voort$^{1}$\thanks{E-mail:
fvdvoort@strw.leidenuniv.nl},
Joop~Schaye$^1$,
Gabriel Altay$^2$,
Tom Theuns$^{2,3}$
\\
$^{1}$Leiden Observatory, Leiden University, Postbus 9513, 2300 RA, Leiden, The Netherlands\\
$^{2}$Institute for Computational Cosmology, Department of Physics, Durham University, South Road, Durham, DH1 3LE, UK \\
$^{3}$Department of Physics, University of Antwerp, Campus Groenenborger, Groenenborgerlaan 171, B-2020, Antwerp, Belgium}

\begin{document}

\date{Accepted not yet. Received \today; in original form \today}

\pagerange{\pageref{firstpage}--\pageref{lastpage}} \pubyear{2011}

\maketitle

\label{firstpage}

\begin{abstract}

Simulations predict that galaxies grow primarily through the accretion of gas that has not gone through an accretion shock near the virial radius and that this cold gas flows towards the central galaxy along dense filaments and streams. There is, however, little observational evidence for the existence of these cold flows. We use a large, cosmological, hydrodynamical simulation that has been post-processed with radiative transfer to study the contribution of cold flows to the observed $z=3$ column density distribution of neutral hydrogen, which our simulation reproduces. We find that nearly all of the H\,\textsc{i} absorption arises in gas that has remained colder than $10^{5.5}$~K, at least while it was extragalactic. In addition, the majority of the H\,\textsc{i} is falling rapidly towards a nearby galaxy, with non-negligible contributions from outflowing and static gas. Above a column density of $N_{\rm H\,\textsc{i}} = 10^{17}~$cm$^{-2}$, most of the absorbers reside inside haloes, but the interstellar medium only dominates for $N_{\rm H\,\textsc{i}} > 10^{21}~$cm$^{-2}$. Haloes with total mass below $10^{10}$~M$_\odot$ dominate the absorption for $10^{17}<N_{\rm H\,\textsc{i}} < 10^{21}$~cm$^{-2}$, but the average halo mass increases sharply for higher column densities. Although very little of the H\,\textsc{i} in absorbers with $N_{\rm H\,\textsc{i}} \lesssim 10^{20}~$cm$^{-2}$ resides inside galaxies, systems with  $N_{\rm H\,\textsc{i}} > 10^{17}~$cm$^{-2}$ are closely related to star formation: most of their H\,\textsc{i} either will become part of the interstellar medium before $z=2$ or has been ejected from a galaxy at $z>3$. Cold accretion flows are critical for the success of our simulation in reproducing the observed rate of incidence of damped Lyman-$\alpha$ and particularly that of Lyman limit systems. We therefore conclude that cold accretion flows exist and have already been detected in the form of high column density H\,\textsc{i} absorbers.

\end{abstract}

\begin{keywords}
galaxies: evolution -- galaxies: formation -- galaxies: high-redshift -- intergalactic medium -- cosmology: theory -- quasars: absorption lines
\end{keywords}

\section{Introduction}

Galaxies grow by accreting gas from their haloes and their haloes grow by accreting gas from the intergalactic medium (IGM). Feedback from star formation and active galactic nuclei returns a significant fraction of the interstellar medium (ISM) to the halo and may even blow it out of the halo into the IGM. This cycle of inflow and outflow makes the circumgalactic medium and the IGM vital ingredients for our understanding of the formation and evolution of galaxies. 

Theoretically we expect that gas accreting onto haloes with sufficiently low circular velocities will not shock-heat to the virial temperature of the halo, but will instead flow in cold (i.e.\ $T \sim 10^4~{\rm K}$) and relatively unimpeded. This so-called ``cold accretion'' will happen if the cooling time of virialized gas is too short to maintain a hot, hydrostatic halo \citep[e.g.][]{White1991, Birnboim2003, Dekel2006}. The existence of such a cold accretion mode has been confirmed by simulations, which have furthermore demonstrated that cold-mode accretion can also be important for haloes sufficiently massive to contain hot, hydrostatic gas. Because gas accretes preferentially along the filaments of the cosmic web, the streams of infalling gas have relatively high gas densities and correspondingly low cooling times. This allows the cold streams to penetrate the hot, hydrostatic haloes surrounding massive galaxies, particularly at high redshift \citep[e.g.][]{Keres2005, Ocvirk2008, Dekel2009a, Keres2009a, Brooks2009, Voort2011a, Voort2011b, VoortSchaye2011, Fumagalli2011, Faucher2011}.  

Although the dilute halo gas that has been shock-heated to the virial temperature is routinely detected in X-ray observations of clusters and groups of galaxies, and has perhaps even been seen around individual galaxies \citep[e.g.][]{Crain2010a, Crain2010b, Anderson2011}, there is no clear, direct observational evidence for cold-mode accretion. However, \citet{Rauch2011} have observed cold, filamentary, infalling gas which could be cold-mode accretion. It has been claimed that the diffuse Lyman-$\alpha$ emission detected around some high-redshift galaxies is due to cold accretion \cite[e.g.][]{Dijkstra2009,Goerdt2010}, but both simulations and observations indicate that the emission is more likely scattered light from central H\,\textsc{ii} regions \cite[e.g.][]{Furlanetto2005,Faucher2010,Steidel2010,Hayes2011,Rauch2011}. Individual H\,\textsc{i} absorbers have also been suggested to be evidence for cold accretion based on their proximity to a galaxy and their low metallicity \cite[]{Ribaudo2011,Giavalisco2011}, but it is difficult to make strong statements for an individual gas cloud, particularly since H\,\textsc{i} near galaxies does not need to be inflowing, even if it has a low metallicity. Given that outflows are routinely detected in the form of blueshifted interstellar absorption lines in the spectra of star-forming galaxies \citep[e.g.][]{Steidel2010, Rubin2010, Rakic2011a}, one may wonder why inflowing gas is not commonly seen in the form of redshifted absorption lines. It is, however, quite possible that the inflowing material has such small cross-sections that the signal is completely swamped by outflowing material \citep[e.g.][]{FaucherKeres2011, Stewart2011a}. 

It is challenging to observe cold gas around distant galaxies in emission, because it typically has densities that are low compared to that of the cold ISM. Absorption line measurements are therefore very important, with the Lyman-$\alpha$ line of neutral hydrogen being the most sensitive probe. Indeed, H\,\textsc{i} absorption can be detected both within and outside of galaxies and at $z\sim 3$ the column density distribution has been measured over ten orders of magnitude in
$N_\mathrm{H\,\textsc{i}}$ \citep[e.g.][]{Tytler1987, Kim2002, Peroux2005, Omeara2007, Prochaska2009, Noterdaeme2009, Prochaska2010}. By correlating the H\,\textsc{i} absorption in the spectra of
background quasars with both the transverse and line of sight separations from foreground galaxies, \citet{Rakic2011b} have recently presented strong evidence for infall of cold gas on scales of $\sim1.4-2.0$ proper Mpc at $z\sim2.4$. 

The low column density ($N_\mathrm{H\,\textsc{i}}<10^{17.2}$~cm$^{-2}$) material is known as the Lyman-$\alpha$ forest and originates mostly in the photo-ionized IGM \citep[e.g.][]{Bi1992, Cen1994, Hernquist1996, Theuns1998, Schaye2001a}. Because these systems are optically thin, they are relatively easy to model in simulations. Lyman Limit Systems (LLSs; $10^{17.2}<N_\mathrm{H\,\textsc{i}}<10^{20.3}$~cm$^{-2}$) and Damped Lyman-$\alpha$ Systems (DLAs; $N_\mathrm{H\,\textsc{i}}>10^{20.3}$~cm$^{-2}$) are optically thick to Lyman limit photons. Because the gas comprising these strong absorbers is partially shielded from the ambient UV radiation, it will be more neutral, and will also be colder, than if it were optically thin. These high column density systems are harder to model, because in order to know which gas is self-shielded, one needs to perform a radiative transfer calculation. Additionally, at the highest gas densities a fraction of the hydrogen will be converted into molecules, reducing the atomic hydrogen fraction\footnote{In this work we mean atomic hydrogen when we write ``neutral hydrogen'' or ``H\,\textsc{i}''.} \citep[e.g.][]{Schaye2001b, Krumholz2009}. 

The difficulty with interpreting absorption-line studies is that they only allow us to study the gas along the line-of-sight direction. It is therefore hard to determine what kind of objects are causing the absorption. Simulations that reproduce the observed absorption-line statistics can be employed to study the three-dimensional distribution of the absorbing gas and to guide the interpretation of the observations.

A lot of progress has been made in understanding and modelling the H\,\textsc{i} column density distribution function \citep[e.g.][]{Katz1996, Schaye2001a, Schaye2001b, Zheng2002, Altay2011, McQuinn2011, Fumagalli2011}. Building on previous work, we ask what kind of gas is causing the H\,\textsc{i} absorption. \citet{Altay2011} have post-processed the reference model of the OWLS suite of cosmological simulations (see Section~\ref{sec:sim}) with radiative transfer and found that the predicted $z=3$ H\,\textsc{i} column density distribution agreed with observations over the full range of observed column densities. We analyse their simulation, selecting different subsets of the gas to see what fraction of the high column density absorption is due to what kind of gas. The selections are based on the maximum past temperature, membership of haloes and the ISM, and the radial velocity towards the nearest galaxy. These properties are all important to determine whether or not the gas is accreting in the cold mode. We find that almost all high column density absorption arises in gas that has never been hotter than $10^{5.5}$~K, at least not while it was extragalactic, and that resides inside of haloes. Most of this gas is strongly linked to star formation because it is either currently in the ISM, has been part of the ISM, or will become part of the ISM before $z=2$. Inflowing gas provides the largest contribution to the H\,\textsc{i} column density distribution function, but outflowing and static gas cannot be neglected.

In Section~\ref{sec:sim} we describe the simulation we used. The different gas samples are listed and described in Section~\ref{sec:masks}. In Section~\ref{sec:HI} we determine the contribution of the different gas samples to the total amount of gas and H\,\textsc{i} and to the H\,\textsc{i} column density distribution function. We discuss our results and conclude in Section~\ref{sec:concl}.

\section{Simulations} \label{sec:sim}

We use a modified version of \textsc{gadget-3} \citep[last described in][]{Springel2005}, a smoothed particle hydrodynamics (SPH) code that uses the entropy formulation of SPH \citep{Springel2002}, which conserves both energy and entropy where appropriate. This work is part of the OverWhelmingly Large Simulations (OWLS) project \citep{Schaye2010}, which consists of a large number of cosmological simulations with varying (subgrid) physics. Here we make use of the so-called `reference' model, with the difference that it was run with updated cosmological parameters. We have tested runs with different supernova feedback and with AGN feedback and we found our conclusions to be robust to changes in the model. The model is fully described in \citet{Schaye2010} and we will only summarize its main properties here.

The simulation assumes a $\Lambda$CDM cosmology with parameters $\Omega_\mathrm{m} = 1 - \Omega_\Lambda = 0.272$, $\Omega_\mathrm{b} = 0.0455$, $h = 0.704$, $\sigma_8 = 0.81$, $n = 0.967$. These values are taken from the WMAP year~7 data \citep{Komatsu2011}. In the present work, we use the simulation output at redshifts 3 and 2.

A cubic volume with periodic boundary conditions is defined, within which the mass is distributed over $512^3$ dark matter and as many gas particles. The box size (i.e.\ the length of a side of the simulation volume) of the simulation used in this work is 25~$h^{-1}$ comoving Mpc. The (initial) particle masses for baryons and dark matter are $2.1\times10^6$~M$_\odot$ and $1.0\times10^7$~M$_\odot$, respectively. 
The gravitational softening length is 2~$h^{-1}\,$comoving kpc, i.e.\ 1/25 of the mean dark matter particle separation, with a maximum of 0.5~$h^{-1}$proper kpc.

Star formation is modelled according to the recipe of \citet{Schaye2008}. A polytropic equation of state $P_\mathrm{tot}\propto\rho_\mathrm{gas}^{4/3}$ is imposed for densities exceeding $n_\mathrm{H}=0.1$~cm$^{-3}$, where $P_\mathrm{tot}$ is the total pressure and $\rho_\mathrm{gas}$ the density of the gas. Gas particles with proper densities $n_\mathrm{H}\ge0.1$~cm$^{-3}$ and temperatures $T\le10^5$~K are moved onto this equation of state and can be converted into star particles. The star formation rate (SFR) per unit mass depends on the gas pressure and reproduces the observed Kennicutt-Schmidt law \citep{Kennicutt1998} by construction.

Feedback from star formation is implemented using the prescription of \citet{Vecchia2008}. About 40 per cent of the energy released by type II supernovae is injected locally in kinetic form, while the rest of the energy is assumed to be lost radiatively. The initial wind velocity is 600~km~s$^{-1}$ and the initial mass loading is two, meaning that, on average, a newly formed star particle kicks twice its own mass in neighbouring gas particles.

\begin{table*} 
\caption{\label{tab:sample} \small List of selection criteria for the various gas samples in the same order as in Figures \ref{fig:bar} and \ref{fig:CDDFfrac}.}
\begin{tabular}[t]{@{}p{0.5cm}@{}p{0.2cm}@{}p{5cm}@{}p{0.2cm}@{}p{11cm}@{}}
\hline
\hline 
\# & & name & & selection \\
\hline
\hline 
1. & & \textit{all}  & & all gas \\
\hline
2. & & \textit{Tmax$<$5p5} & & gas with $T_\mathrm{max}\le10^{5.5}$~K \\
\hline
3. & & \textit{inhalo} & & halo gas (with $M_\mathrm{halo}\ge10^9$~M$_\odot$) \\
4. & & \textit{inmainhalo} & & gas in main haloes (with $M_\mathrm{main\ halo}\ge10^9$~M$_\odot$) \\
5. & & \textit{insubhalo} & & gas in satellites (with $M_\mathrm{satellite}\ge10^8$~M$_\odot$) \\
\hline
6. & & \textit{inhalo9to10} & & gas in haloes with $10^9\le M_\mathrm{halo}<10^{10}$~M$_\odot$ \\
7. & & \textit{inhalo10to11} & & gas in haloes with $10^{10}\le M_\mathrm{halo}<10^{11}$~M$_\odot$ \\
8. & & \textit{inhalo11to12} & & gas in haloes with $10^{11}\le M_\mathrm{halo}<10^{12}$~M$_\odot$ \\
9. & & \textit{inhalo12to13} & & gas in haloes with $10^{12}\le M_\mathrm{halo}<10^{13}$~M$_\odot$ \\
\hline
10. & & \textit{inflowing} & & gas that is moving towards the nearest (nearest in units of $R_\mathrm{vir}$) halo centre with a peculiar velocity greater than a quarter of its circular velocity \\ 
11. & & \textit{static} & & gas that is neither inflowing (\#10) nor outflowing (\#12) \\ 
12. & & \textit{outflowing} & & gas that is moving away from the nearest (nearest in units of $R_\mathrm{vir}$) halo centre with a peculiar velocity greater than a quarter of its circular velocity \\ 
\hline
13. & & \textit{inhalo\_inflowing} & & halo gas that is moving towards the halo centre with a peculiar velocity greater than a quarter of its circular velocity \\
14. & & \textit{inhalo\_static} & & halo gas that is neither inflowing (\#10) nor outflowing (\#12) \\ 
15. & & \textit{inhalo\_outflowing} & & halo gas that is moving away from the halo centre with a peculiar velocity greater than a quarter of its circular velocity \\
\hline
16. & & \textit{inISM} & & gas which is in the ISM at $z=3$ \\
17. & & \textit{pastISM} & & gas which is not in the ISM at $z=3$, but was at some redshift $z>3$ \\
18. & & \textit{futureISM} & & gas which is not in the ISM at $z=3$, but will be before $z=2$.\\
19. & & \textit{futurefirstISM} & & gas which is not in the ISM at $z\ge3$, but will be before $z=2$. \\
\hline
\hline
\end{tabular}
\end{table*}

The abundances of eleven elements released by massive stars and intermediate mass stars are followed as described in \citet{Wiersma2009b}. We assume the stellar initial mass function (IMF) of \citet{Chabrier2003}, ranging from 0.1 to 100~M$_\odot$. As described in \citet{Wiersma2009a}, radiative cooling and heating are computed element by element in the presence of the cosmic microwave background radiation and the \citet{Haardt2001} model for the UV/X-ray background from galaxies and quasars in the optically thin limit.

We apply a self-shielding correction in post-processing by using the radiative transfer calculations described in \citet{Altay2011} and Altay et al. (in preparation). This is an iterative method that uses reverse ray-tracing (i.e.\ shooting rays starting from the gas element) to compute the optical depth and to calculate a new neutral fraction until the results converge. The method assumes that the radiation field is dominated by the UV background radiation. \citet{Altay2011} have shown that this provides a good match to the observed H\,\textsc{i} column density distribution function. The temperature of the ISM gas in the simulation is determined by the imposed effective pressure and therefore not realistic, so we set it to $10^4$~K, typical of the warm neutral medium. Gas with lower densities ($n_\mathrm{H}<0.1$~cm$^{-3}$) was assumed to be optically thin during the simulation, whereas this may not be the case because of self-shielding. We use the temperatures that were calculated in the presence of the UV background to be self-consistent with the simulation, even though the actual gas temperatures will be lower in the self-shielded case. These choices do not affect the results described here. Hydrogen is converted into molecules with a prescription based on observations by \citet{Blitz2006}, who provide the molecular fraction as a function of the gas pressure.

Recently \citet{Keres2011} have shown that the sizes of galaxy discs depend on the computational method used (but see also \citet{Hummels2011}). However,
\citet{Sales2010} have shown that changes in the feedback model lead to dramatic variations in the morphology of discs, as well as to variations of up to an order of magnitude in disc masses. Indeed, \citet{Scannapieco2011} have shown explicitly that variations in the feedback prescription are much more important than the numerical scheme used to solve the hydrodynamics. To assess the robustness of our conclustions, we have redone our analysis using simulations with different feedback prescriptions, but do not show the results. The simulations reproduce the observed H\,\textsc{i} column density distribution function independent of the feedback scheme (Altay et al. in preparation). Although the relative contributions of the different gas selections to the H\,\textsc{i} absorption vary somewhat between different simulations, the general conclusions are unchanged. We therefore conclude that the uncertainties associated with the feedback and the numerical method are unlikely to be important for our conclusions.

\section{Gas samples} \label{sec:masks}

We compute the $z=3$ contribution of gas with different properties to the H\,\textsc{i} column density distribution function. The gas selections that we consider are listed in Table~\ref{tab:sample}. They are based on the maximum past temperature, $T_\mathrm{max}$, membership of haloes, the peculiar velocity towards the nearest halo, $v_\mathrm{rad}$, and on whether the gas is part of the ISM. These properties are all important to determine whether or not this gas is accreting in the cold mode. Cold-mode gas should have a low temperature at least until it becomes part of the ISM. Gas has only been accreted if it is inside a halo. It is likely to accrete onto a halo or its central galaxy if its radial peculiar velocity towards the halo centre is high, but note that gas orbiting around the central galaxy can also be built up by cold flows and could be viewed as evidence for cold accretion \citep{Stewart2011b}. The gas is certainly accreting onto a galaxy if it enters the ISM of a galaxy and becomes star forming in the near future. Below, we describe the selections in detail.

The maximum past temperature, $T_\mathrm{max}$, is the maximum value that the temperature of a gas element has reached over the entire past simulation history. High time resolution is achieved by storing a separate variable during the simulation which is updated every time-step if the gas reaches a higher temperature than the previous value of $T_\mathrm{max}$. The artificial temperature that a gas particle has while it is on the imposed equation of state, and thus star forming, is ignored. Because we are interested in the temperature history before the gas accretes onto a galaxy, this is consistent with our aims. Cold-mode gas does not experience a shock at the virial radius and reaches a maximum temperature of $\sim10^5$~K just before accretion onto a halo due to heating by the UV background \citep{Voort2011a}. We therefore select cold-mode gas by using $T_\mathrm{max}\le10^{5.5}$~K (sample \textit{Tmax$<$5p5} of Table~\ref{tab:sample}). While hot-mode gas has by definition been hotter than $10^{5.5}$~K, its \textit{current} temperature can be much lower \citep[e.g.][]{VoortSchaye2011}.

We note that $T_{\rm max}$ may underestimate the true maximum past temperature. In SPH simulations a shock is smeared out over a few smoothing lengths, leading to in-shock cooling \citep{Hutchings2000}. If the cooling time is of the order of or smaller than the time step, then the maximum temperature will be underestimated. \citet{Creasey2011} have shown that a particle mass of $10^6$~$M_\odot$ is sufficient to avoid numerical overcooling of accretion shocks onto haloes, which is very close to the resolution used in this simulation. Interpolating their results indicates that our mass resolution is indeed sufficient. Even with infinite resolution, the post-shock temperatures may, however, not be well defined. Because they have different masses, electrons and protons will have different temperatures in the post-shock gas and it may take some time before they equilibrate
through collisions or plasma effects. We have ignored this complication. Another effect, which was also not included in our simulation, is that shocks may be preceded by the radiation from the shock, which may affect the temperature evolution. Ignoring these issues, \citet{Voort2011a} showed that the distribution of $T_\mathrm{max}$ is bimodal and that a cut at $T_{\rm max}=10^{5.5}$~K naturally divides the gas into filamentary cold- and diffuse hot-mode accretion and that it produces similar results as studies based on adaptive mesh refinement simulations. 

Haloes are found by using a Friends-of-Friends (FoF) algorithm. If the separation between two dark matter particles is less than 20 per cent of the average separation (the linking length $b = 0.2$), they are placed in the same group. Assuming a radial density profile $\rho (r)\propto r^{-2}$, corresponding to a flat rotation curve, such a group has an average overdensity of $\langle\rho_\mathrm{halo}\rangle/\langle\rho\rangle\approx180$ \citep[e.g.][]{Lacey1994}, which is close to the value for a virialized object predicted by the top-hat spherical collapse model. Baryonic particles are placed in a halo if their nearest dark matter neighbour is part of the halo. Haloes can contain both bound and unbound particles. The selections are made by using haloes with total mass $M_\mathrm{halo}\ge10^9$~M$_\odot$, corresponding to approximately 100 dark matter particles. A gas particle is considered to be in a halo if it is a member of a Friends-of-Friends group (\textit{inhalo}).

We use \textsc{subfind} \citep{Springel2001, Dolag2009} on the FoF output to identify gravitationally bound subhaloes. The main halo is the most massive subhalo in a FoF halo and the remaining subhaloes are classified as satellites. Because only bound particles are included in this definition, some unbound particles are not attached to any subhalo, although they are part of the FoF halo. The gas is in a main halo if it is a member of the most massive subhalo, with $M_\mathrm{main\ halo}\ge10^9$~M$_\odot$ (\textit{inmainhalo}) and the gas is in a satellite if it is a member of a subhalo, with $M_\mathrm{satellite}\ge10^8$~M$_\odot$, that is not a main halo (\textit{insubhalo}). 

The halo gas (\textit{inhalo}) is subdivided into four mass bins, $10^9\le M_\mathrm{halo}<10^{10}$~M$_\odot$ (\textit{inhalo9to10}), $10^{10}\le M_\mathrm{halo}<10^{11}$~M$_\odot$ (\textit{inhalo10to11}), $10^{11}\le M_\mathrm{halo}<10^{12}$~M$_\odot$ (\textit{inhalo11to12}), and $10^{12}\le M_\mathrm{halo}<10^{13}$~M$_\odot$ (\textit{inhalo12to13}), containing 38\,504, 3626, 275, and 17 haloes, respectively.

\begin{figure*}
\center
\includegraphics[scale=0.6]{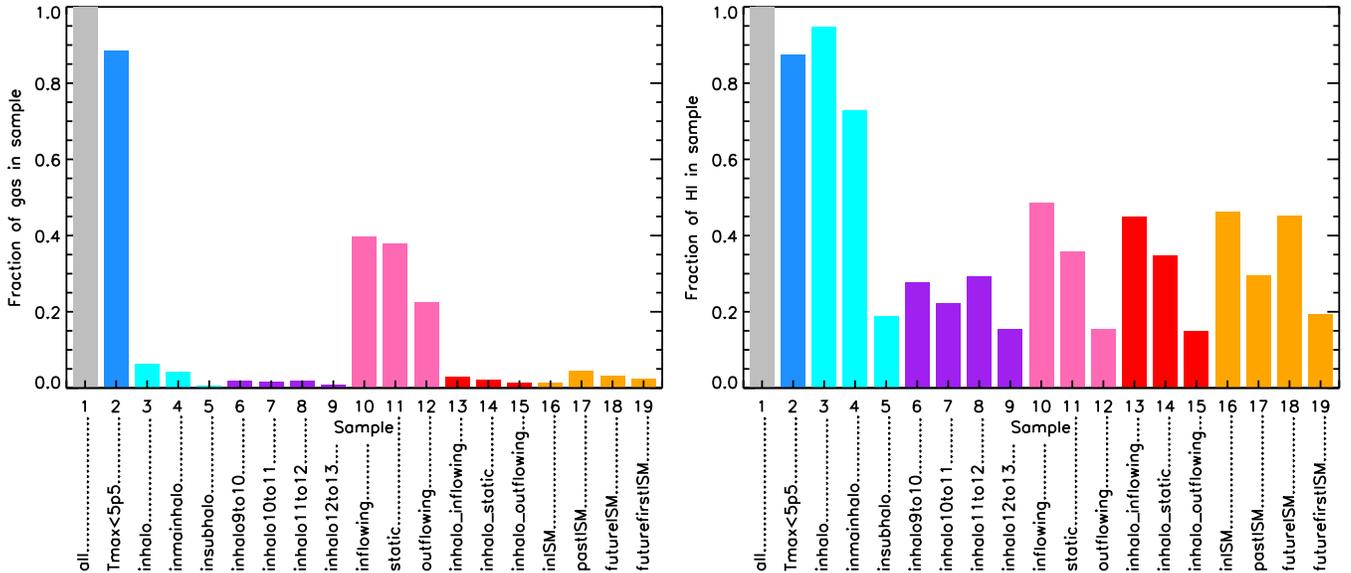}
\caption {\label{fig:bar} Fraction of the total amount of gas (left) and H\,\textsc{i} (right) at $z=3$ accounted for by the selections listed in Table~\ref{tab:sample}. Many samples (\#3--9 and \#13--19) contain a much larger fraction of the H\,\textsc{i} mass than of the total gas mass.}
\end{figure*}

The radial peculiar velocity of every gas particle is calculated with respect to the nearest halo. The position (and the centre) of the halo is defined as the position of its most bound particle. The virial radius, $R_\mathrm{vir}$, is the radius around this centre which encloses an average density that agrees with the prediction from top-hat spherical collapse calculations in a $\Lambda$CDM cosmology \citep{Bryan1998}. Distances towards a halo, $d$, are normalized by $R_\mathrm{vir}$ of that halo. The nearest halo is the one for which $d/R_\mathrm{vir}$ is minimum. The halo's peculiar velocity is defined to be the centre of mass velocity of its main subhalo and is subtracted from the gas peculiar velocity to determine the relative radial peculiar velocity. If the gas is located inside a halo, its radial velocity is calculated with respect to the centre of that halo, even if it is also in a satellite. The selections are made by comparing the radial velocity of the gas to a quarter of the circular velocity, $v_\mathrm{circ}=\sqrt{GM_\mathrm{vir}/R_\mathrm{vir}}$, of the nearest halo, where $M_{\rm vir}$ is the total mass inside $R_{\rm vir}$. The gas is inflowing when $v_\mathrm{rad}\le-0.25v_\mathrm{circ}$ (\textit{inflowing}), static when $-0.25v_\mathrm{circ}<v_\mathrm{rad}<0.25v_\mathrm{circ}$ (\textit{static}), and outflowing when $v_\mathrm{rad}\ge 0.25v_\mathrm{circ}$ (\textit{outflowing}). Note that gas near or inside a small halo embedded in a filament may be classified as outflowing with respect to its nearest halo, whereas it might be inflowing with respect to a much larger halo that is being fed by this filament. It is hard to unambiguously identify this gas, so we just caution the reader that the amount of inflowing gas may be an underestimate. This would only strengthen our conclusions.

The definition of the nearest halo is somewhat arbitrary and the gas may be many virial radii away and unrelated to the nearest halo. This is why we make the same radial velocity cuts for all gas (samples 10--12 of Table~\ref{tab:sample}) also for gas that is already inside a (FoF) halo, so that we know it has already been accreted by the halo (samples 13--15 of Table~\ref{tab:sample}). Halo gas which is inflowing is in the process of accreting onto the central galaxy (\textit{inhalo\_inflowing}).

In our simulation gas becomes star forming when its density exceeds $n_\mathrm{H}=0.1$~cm$^{-3}$ while its temperature $T\le10^5$~K. Because of supernova feedback, the gas can leave the ISM and re-enter it \citep[e.g.][]{Oppenheimer2010}. We will call this recycled gas even though it may not have been part of a star before it was ejected from the ISM. We keep track of the time a gas particle last reached or left the ISM. We select all the gas that is star forming at $z=3$, i.e.\ the ISM (\textit{inISM}), star forming before but not at $z=3$, i.e.\ recycled gas (\textit{pastISM}), star forming before $z=2$ but not at $z=3$, i.e.\ gas that will accrete onto a galaxy before $z=2$ (\textit{futureISM}), or star forming before $z=2$ but not at $z\ge3$, i.e.\ gas that will accrete onto the galaxy before $z=2$ for the first time (\textit{futurefirstISM}). The time elapsed between $z=3$ and $z=2$ is about 1.2~Gyr, which is similar to the gas consumption time scale implied by the observed Kennicutt-Schmidt star formation law for typical ISM densities and indicates whether the gas will accrete onto a galaxy in the near future.

\section{Results} \label{sec:HI}

\subsection{Gas and H\,{\sc i} fractions}

Many of the gas samples listed in Table~\ref{tab:sample} contain a much larger fraction of the total H\,\textsc{i} mass than of the total gas mass. This is illustrated in Figure~\ref{fig:bar}, which shows the fraction of the total amount of gas (left panel) and the fraction of the total amount of neutral hydrogen (right panel) accounted for by the samples listed in Table~\ref{tab:sample}. The density parameter of all the gas $\Omega_\mathrm{gas}=0.15787$ and of neutral hydrogen $\Omega_\mathrm{H\,\textsc{i}}=0.00178$, so at $z=3$ only about one per cent of the gas mass consists of neural hydrogen. Selections based on halo membership (samples 3--9 and 13--15) and on participation in star formation (samples 16--19) account for little of the gas mass, but a significant part of the H\,\textsc{i} mass.

The maximum past temperature selection (\textit{Tmax$<$5p5}) contains almost all gas in the simulation. This is not so surprising, because most of the mass in the universe is in the IGM and the UV background can only heat this gas to temperatures $\lesssim10^5$~K. The same selection also contains almost all neutral hydrogen gas. Gas that shocked to $T>10^{5.5}$~K while it was extragalactic, either through accretion shocks or shocks caused by galactic winds, accounts for only about 10 per cent of both the gas and H\,\textsc{i}. This already tells us that \emph{most gas and most neutral hydrogen is cold-mode gas}. 

Note that this is not a trivial result. While it is true that gas with temperatures above $10^{5.5}$~K will be collisionally ionized and therefore difficult to see in H\,\textsc{i} absorption, we distinguish cold- and hot-mode gas based on the maximum \textit{past} temperature rather than the \textit{current} temperature. Gas parcels accreted in the hot mode can cool down, reach temperatures similar to those of cold-mode gas, and show up in H\,\textsc{i} absorption.

Haloes contain less than 10 per cent of the gas, but more than 90 per cent of the neutral hydrogen (\textit{inhalo}). Gas outside haloes contributes little to the H\,\textsc{i} content, because the neutral fraction increases with density and the highest densities are found inside haloes. Combining this with the finding that sample \textit{Tmax$<$5p5} accounts for nearly 90 per cent of the neutral hydrogen, proves that most H\,\textsc{i} absorption arises from gas that has been accreted onto haloes in the cold-mode. All main haloes contain seven times more gas and four times more H\,\textsc{i} than all satellites (compare \textit{inmainhalo} and \textit{insubhalo}).

Haloes from the entire mass range contain a significant part of the H\,\textsc{i} mass (samples 6--9). Haloes in the highest mass bin are very rare and their total contribution is the smallest (15 per cent; \textit{inhalo12to13}). Haloes below $M_\mathrm{halo}=10^9$~M$_\odot$, and therefore below our resolution limit, are not expected to contain a large amount of H\,\textsc{i}, because we already account for the vast majority of H\,\textsc{i} with higher mass haloes (\textit{inhalo}) and our simulations reproduce the observed H\,\textsc{i} distribution \citep{Altay2011}.

Approximately 40 per cent of the gas and half of the neutral hydrogen is inflowing faster than a quarter of the circular velocity of the nearest (nearest in units of $R_\mathrm{vir}$) halo (\textit{inflowing}). The contribution is even larger for all inflowing gas, without a minimum radial velocity threshold (about 65 per cent; not shown). Outflows always account for less than 25 per cent of the gas and H\,\textsc{i} mass (\textit{outflowing}). The contribution of outflowing gas and of static gas (i.e.\ gas moving either in or out at velocities below $0.25v_\mathrm{circ}$; \textit{static}) is lower for H\,\textsc{i} than for all gas, whereas the contribution of inflowing gas is higher.

Within haloes, the importance of inflows (\textit{inhalo\_inflowing}), with respect to outflows (\textit{inhalo\_outflowing}) and static gas (\textit{inhalo\_static}), increases slightly. This is not visible in Figure \ref{fig:bar}, because the bars are too small.
The H\,\textsc{i} inside haloes is distributed over the inflowing, static, and outflowing components in the same way as all H\,\textsc{i} (i.e.\ both inside and outside of haloes), because haloes account for nearly all the H\,\textsc{i}. Thus, about half of the H\,\textsc{i} can be accounted for by gas inside the virial radius that is falling towards the halo centre with a velocity greater than 25 per cent of the halo's circular velocity. 

Very little gas is contained in the ISM of galaxies, but almost half of the H\,\textsc{i} mass is (\textit{inISM}). The amount of gas that has been ejected from galaxies at $z>3$ (\textit{pastISM}) is greater than the amount that is in the ISM at $z=3$ (\textit{inISM}), but the situation is reversed for H\,\textsc{i}. Similarly, more gas has been ejected from galaxies at $z>3$ (\textit{pastISM}) than will accrete onto galaxies at $2\le z<3$ (\textit{futureISM}), but it is the other way around for neutral hydrogen. Taken together, gas that is currently in the ISM and gas that will enter the ISM within 1.2~Gyr (\textit{inISM} + \textit{futureISM}) account for most of the H\,\textsc{i} in the universe. Gas/H\,\textsc{i} that will accrete onto a galaxy at $2\le z<3$ \textit{for the first time} (\textit{futurefirstISM}) accounts for most/less than half of the total amount of gas/H\,\textsc{i} that will accrete at $2\le z<3$ (\textit{futureISM}).

\subsection{Spatial distribution: A visual impression}

\begin{figure*}
\center
\includegraphics[scale=1]{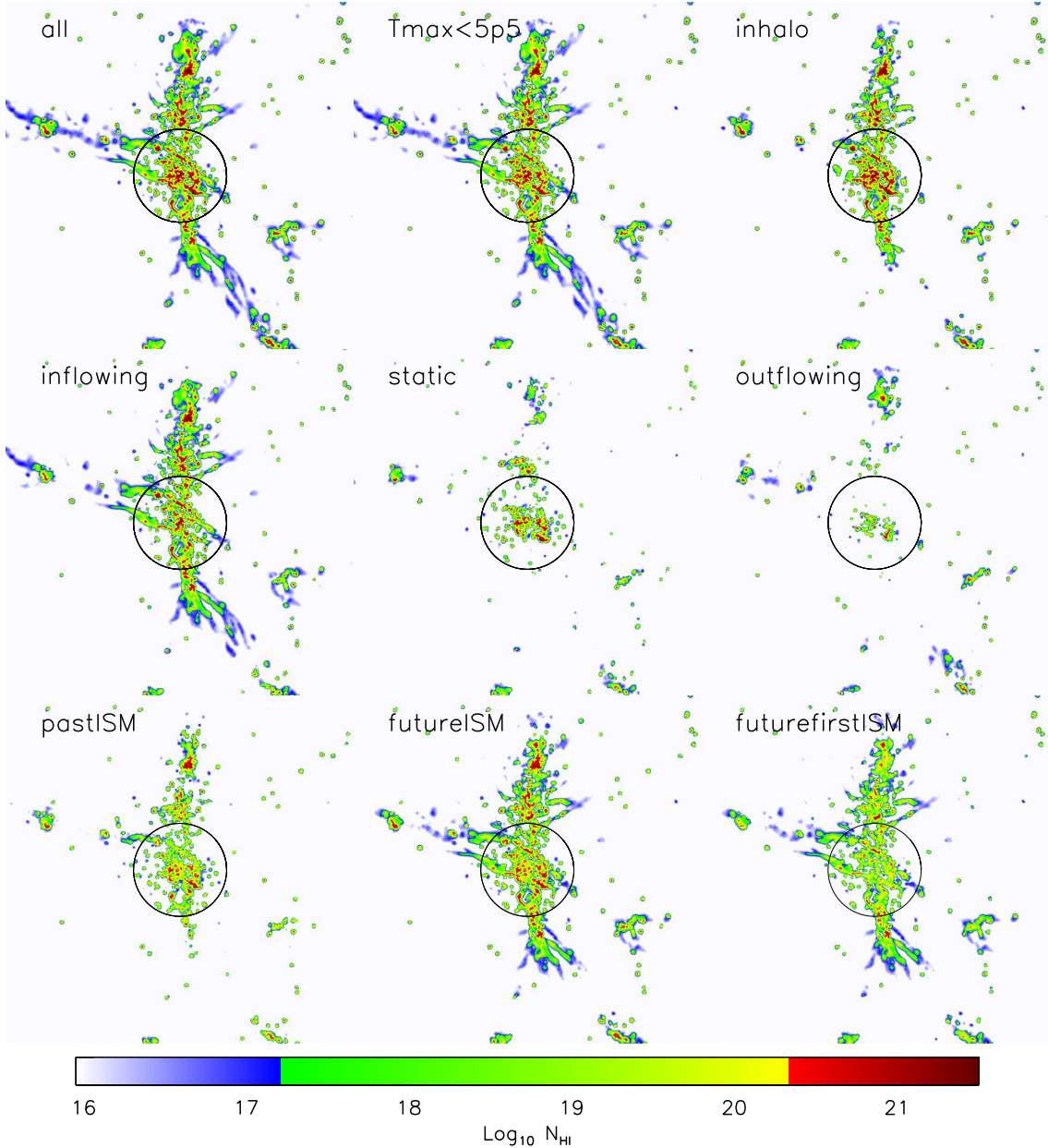}
\caption {\label{fig:image} H\,\textsc{i} column density images for a cubic region of 2 comoving~$h^{-1}\,$Mpc on a side centred on a $10^{12.4}$~M$_\odot$ halo at $z=3$, one of the most massive haloes in this simulation. For reference, the virial radius of this halo is 268~comoving~$h^{-1}\,$kpc and indicated with black circles. For all but the first panel, the H\,\textsc{i} column densities were computed including only gas selected to be part of the samples indicated in each panel and listed in Table~\ref{tab:sample}. Some gas samples trace the full H\,\textsc{i} morphology, whereas others only trace a subset of the high column density sightlines.}
\end{figure*}

To gain insight into the spatial distribution of the neutral hydrogen, Figure~\ref{fig:image} shows the H\,\textsc{i} column densities at $z=3$ in a 2~by~2~comoving~$h^{-1}\,$Mpc region centred on a $10^{12.4}$~M$_\odot$ halo, one of the most massive objects in the simulation. The Figure is meant to illustrate the spatial distribution of the H\,\textsc{i} corresponding to some of the samples listed in Table~\ref{tab:sample}. Some gas samples trace only a subset of the high column density sightlines, while others trace the full morphology of the H\,\textsc{i}.

In the top row, the column densities are shown when including all gas (left), only gas with $T_\mathrm{max}\le10^{5.5}$~K (middle), and only gas inside haloes (right). The vast majority of neutral gas inside and outside haloes satisfies $T_\mathrm{max}\le10^{5.5}$~K, so even for the $10^{12.4}$~M$_\odot$ halo with $T_\mathrm{vir}=10^{6.2}$~K, most of the H\,\textsc{i} is contributed by gas that has $T_\mathrm{max}\le10^{5.5}$ and that has thus not gone through an accretion shock near the virial radius. When we exclude gas outside haloes (top right panel), we lose the filamentary material connecting different haloes, as well as most of the $N_\mathrm{H\,\textsc{i}}<10^{17}$~cm$^{-2}$ sightlines. 

The middle row shows the column density when including only gas that is inflowing towards the nearest (nearest in units of $R_{\rm vir}$) halo faster than a quarter of the circular velocity of that halo (left), only gas outflowing faster than a quarter of the circular velocity (right), and the remaining and hence ``static'' gas (middle). The morphologies of the H\,\textsc{i} gas are very different. The inflowing gas traces most of the extended and filamentary structure of the total H\,\textsc{i} distribution, while the static and outflowing gas are concentrated in the centres of the haloes.

The images shown in the bottom row include only gas that is not part of the ISM at $z=3$. Additionally, the panels include only gas that was part of the ISM for some redshift $z>3$ (i.e.\ ejected gas; left panel), only gas that becomes part of the ISM at $2\le z<3$ (middle), and only gas that joins the ISM at $2\le z<3$ for the first time (right). The ejected gas that is not currently part of the ISM (bottom left panel) is located relatively close to the centres of haloes, where star formation is taking place, but it is more spread out than the actual ISM (not shown). Comparing it to the middle row, with inflowing (left), static (centre), and outflowing (right) gas, we immediately see that some of the ejected gas is already inflowing and thus recycling. Note that it may have been ejected from a different halo than the one it is in now. Most of the lower column density filamentary gas, especially the gas outside the halo, is missing. Even though ejected gas contributes significantly to $N_\mathrm{H\,\textsc{i}}\gtrsim10^{18}$~cm$^{-2}$ gas, it is unimportant for lower column densities. 

If we include all the gas that will reach the ISM between $2\le z<3$, i.e.\ within 1.2~Gyr from the redshift at which we are studying the H\,\textsc{i} column densities (bottom middle panel), we leave the morphology of the $N_\mathrm{H\,\textsc{i}}>10^{16}$~cm$^{-2}$ sightlines intact, including the filaments in between haloes, although they become somewhat weaker. Most, but not all, H\,\textsc{i} gas in filaments close to a halo will accrete onto a galaxy before $z=2$. 

The last panel shows a subset of the gas in the previous panel, because it only includes gas that reaches the ISM between $2\le z<3$ for the first time. The general morphology of H\,\textsc{i} is the same, apart from some extra holes in the central region of the halo, which are dominated by ejected gas. The column densities above $10^{18}$~cm$^{-2}$ are lower for a given pixel.

We conclude that the large-scale H\,\textsc{i} filaments that feed massive haloes consist mostly of gas that has never gone through a virial shock, that is inflowing and that will join the ISM, and hence participate in star formation, in the near future. This gas typically has H\,\textsc{i} column densities similar to those of LLSs or higher for embedded clumps. Within haloes, the infalling cold-mode gas has higher H\,\textsc{i} column densities and tends to have a more clumpy morphology than the H\,\textsc{i} in filaments outside haloes.

\subsection{The column density distribution function}

\begin{figure}
\center
\includegraphics[scale=0.5]{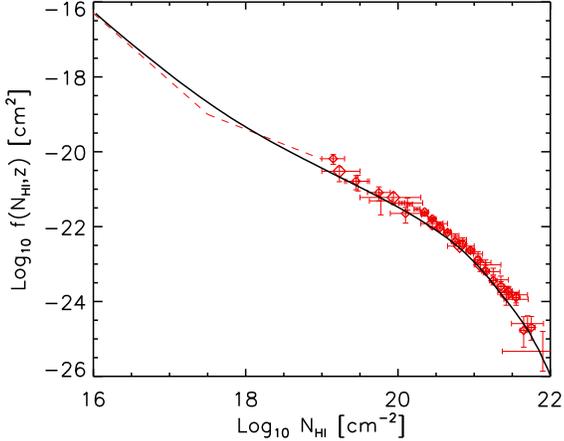}
\caption {\label{fig:CDDF} Predicted H\,\textsc{i} column density distribution function at $z=3$ (black curve) and a compilation of observations around the same redshift \citep[data points:][]{Peroux2005, Omeara2007, Noterdaeme2009, Prochaska2010} and power-law constraints \citep[dashed curve:][]{Prochaska2009}. The black curve is obtained by collapsing the full three-dimensional H\,\textsc{i} distribution onto a plane. The observations are matched well \citep[see also][]{Altay2011}.}
\end{figure}

The H\,\textsc{i} column density distribution function is defined as the number of absorption lines $\mathcal{N}$, per unit column density $dN_\mathrm{H\,\textsc{i}}$, per unit absorption distance $dX$. 
\begin{equation}
f(N_\mathrm{H\,\textsc{i}},z)\equiv\dfrac{d^2\mathcal{N}}{dN_\mathrm{H\,\textsc{i}}dX}\equiv\dfrac{d^2\mathcal{N}}{dN_\mathrm{H\,\textsc{i}}dz}\dfrac{dz}{dX}
\end{equation}
Absorbers with a fixed proper size and a constant comoving number density are distributed uniformly, in a statistical sense, per unit absorption distance along a line-of-sight \citep{Bahcall1969}. The absorption distance is related to the redshift path $dz$ as $dX/dz = H_0(1+z)^2/H(z)$, where $H(z)$ is the Hubble parameter.

The predicted H\,\textsc{i} column density distribution function is shown in Figure~\ref{fig:CDDF} as the black curve. It was calculated by projecting the full three-dimensional gas distribution onto a two-dimensional grid with 16\,384$^2$ pixels, which is sufficient to achieve convergence. Observationally determined power-law slopes at $N_\mathrm{H\,\textsc{i}}<10^{19}$~cm$^{-2}$ \citep{Prochaska2009} and data points at $N_\mathrm{H\,\textsc{i}}>10^{19}$~cm$^{-2}$ \citep{Peroux2005, Omeara2007, Noterdaeme2009, Prochaska2010} have been converted to the cosmology assumed in our simulation and are shown as red, dashed lines and red data points, respectively. The observations of the H\,\textsc{i} column density distribution function are matched well, also at column densities lower than shown here \citep{Altay2011}. The match would have been even better if we had corrected the gas temperatures for the effect of self-shielding \citep[see][]{Altay2011}. Including the radiation from local sources might worsen the agreement, although \citet{Nagamine2010} have shown this effect to be unimportant.

For $N_\mathrm{H\,\textsc{i}}\gtrsim10^{17}$~cm$^{-2}$ nearly all sightlines are dominated by a single absorption system, because they are very rare \citep{Prochaska2010, Altay2011}. For column densities $\lesssim 10^{16}$~cm$^{-2}$ our method of projecting the entire 25 comoving~$h^{-1}\,$Mpc simulation box does not recover the true column density distribution due to the projection of unrelated lower column density absorbers. \citet{Altay2011} therefore determine the column density distribution in this regime by decomposing H\,\textsc{i} absorption spectra into Voigt profiles. Above $N_\mathrm{H\,\textsc{i}}\sim10^{18}$~cm$^{-2}$ the column density distribution function flattens, because self-shielding becomes important \citep{Katz1996, Zheng2002, Altay2011}. It steepens again at $N_\mathrm{H\,\textsc{i}}\gtrsim10^{20.3}$~cm$^{-2}$, because the neutral hydrogen fraction saturates \citep{Altay2011}, and at $N_\mathrm{H\,\textsc{i}}\gtrsim10^{21.5}$~m$^{-2}$ due to the formation of molecules \citep{Schaye2001b, Zwaan2006, Krumholz2009, Noterdaeme2009}.

\begin{figure*}
\center
\includegraphics[scale=0.6]{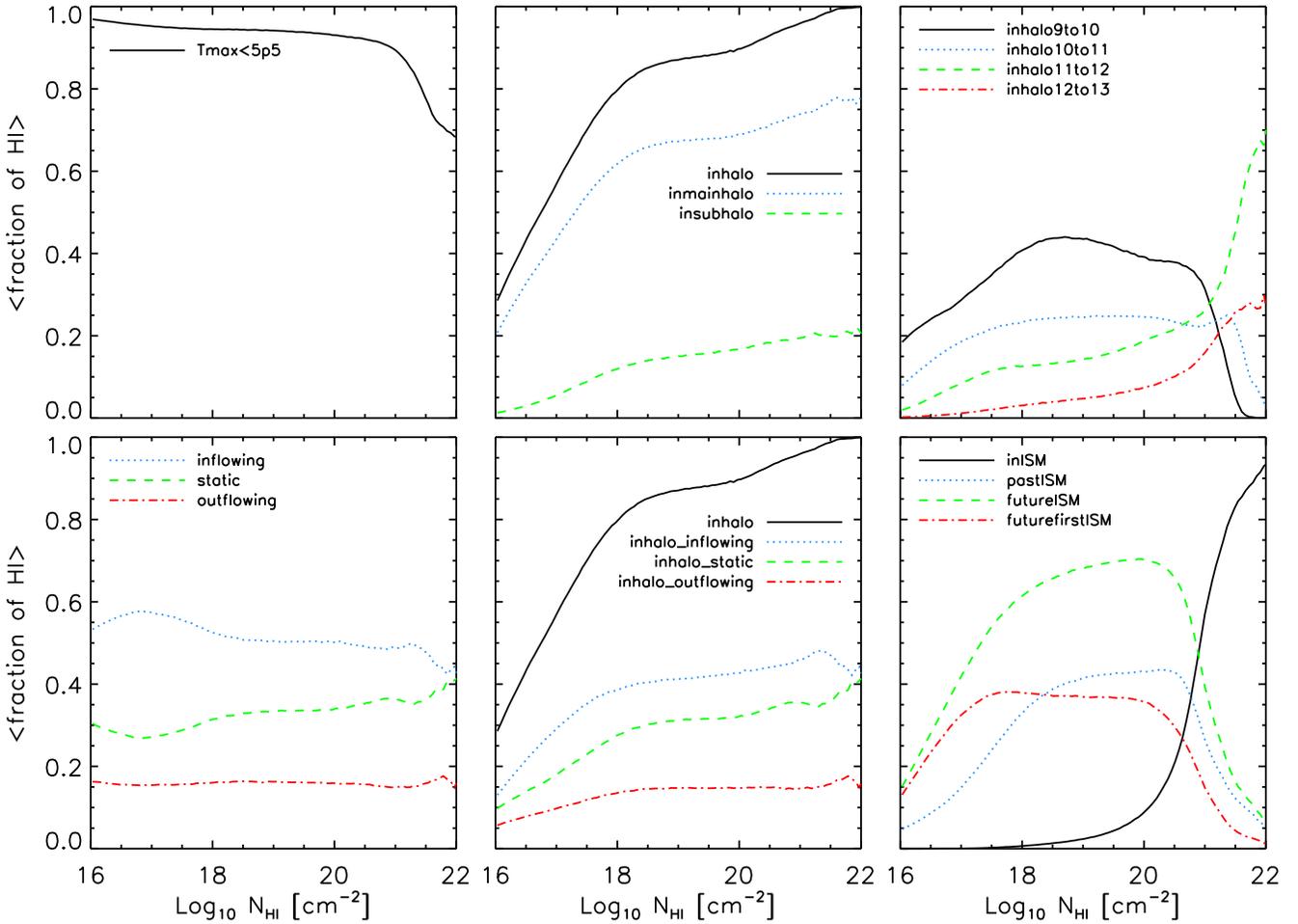}
\caption {\label{fig:CDDFfrac} Mean fraction of the H\,\textsc{i} column density at $z=3$ contributed by the gas which fulfils the selection criteria listed in Table~\ref{tab:sample} as a function of the total H\,\textsc{i} column density. Almost all of the contributing gas has $T_\mathrm{max}<10^{5.5}$~K although hot-mode gas becomes significant, accounting for up to 30 per cent of the absorption, above $10^{21}$~cm$^{-2}$. More than 80 per cent of $N_\mathrm{H\,\textsc{i}}>10^{18}$~cm$^{-2}$ gas is located inside (main) haloes. About 55 per cent of the H\,\textsc{i} is inflowing, but the contributions from static and outflowing gas are about 30 and 15 per cent, respectively, and therefore not negligible. The contribution from gas accreting onto galaxies, i.e.\ gas that will join the ISM in the near future, is dominant for $10^{17} - 10^{21}~$cm$^{-2}$, accounting for up to 70 per cent. Strong DLAs ($N_\mathrm{H\,\textsc{i}}>10^{21}$~cm$^{-2}$) are dominated by ISM gas (contributing more than 60 per cent to the absorption) and the Lyman-$\alpha$ forest by gas that will remain intergalactic down to at least $z=2$ (contributing more than 60 per cent to the absorption).}
\end{figure*}

We can quantify the contribution of different gas samples by calculating the mean fraction of the H\,\textsc{i} column density that is due to each gas selection as a function of the total H\,\textsc{i} column density. This is shown in Figure~\ref{fig:CDDFfrac} for all the samples listed in Table~\ref{tab:sample}. Note that the mean fraction would be 0.5 for a column density $N'_{\rm H\,\textsc{i}}$ if the selection includes all $N'_{\rm H\,\textsc{i}}$ absorbers, but only half of their H\,\textsc{i}. However, a mean fraction of 0.5 would also be obtained if the selection includes half of the absorbers with column density $N'_{\rm H\,\textsc{i}}$, but all the H\,\textsc{i} of the selected absorbers. Samples based on cuts in halo mass are best described by the second case, but most other samples combine the two types of possibilities. 

Figure~\ref{fig:CDDFfrac} shows that most of the H\,\textsc{i} absorption is due to cold-mode gas that is flowing towards the centre of the nearest halo. Above $N_\mathrm{H\,\textsc{i}}=10^{17}$~cm$^{-2}$ this gas is primarily inside haloes. Most of the H\,\textsc{i} gas at these column densities is currently in the ISM or will become part of the ISM within the next 1.2 Gyr. We discuss this in more detail below. 

For absorbers with $N_\mathrm{H\,\textsc{i}}<10^{21}$~cm$^{-2}$ more than 90 per cent of the H\,\textsc{i} column density is due to gas that has never gone through a virial shock, i.e.\ cold-mode gas (top left panel). For higher column densities, which arise mostly in the ISM of galaxies residing in haloes with $M_\mathrm{halo}>10^{11}$~M$_\odot$, this cold-mode fraction drops to 70 per cent. We do not know what caused the remaining gas to be heated above $10^{5.5}$~K (note that the same gas had a much smaller neutral fraction at the time when it was this hot). It could have been a virial shock, an accretion shock at a smaller radius, or a shock associated with a galactic wind. Note that the gas that has $T_{\rm max} > 10^{5.5}$~K because it has been shocked in an outflow, may have been accreted in the cold mode.

Most high column density gas resides in haloes, but the halo fraction drops below 50 per cent for $N_\mathrm{H\,\textsc{i}}<10^{17}$~cm$^{-2}$ (top middle panel). Combining this with the fact that most gas has $T_{\rm max} < 10^{5.5}$~K, we conclude that most LLSs and most DLAs arise from gas inside haloes that has been accreted in the cold mode. Most of the H\,\textsc{i} absorption occurs in main haloes. Satellites contribute less than 20 per cent for all column densities, but their contribution increases slightly with the H\,\textsc{i} column density.

Very low-mass haloes ($10^9<M_\mathrm{halo}<10^{10}$~M$_\odot$) are most abundant and therefore contain the largest amount of H\,\textsc{i} below $N_\mathrm{H\,\textsc{i}}=10^{21}$~cm$^{-2}$, accounting for about 40 per cent of the total absorption (top right panel). For $N_\mathrm{H\,\textsc{i}}>10^{21}$~cm$^{-2}$, a single very low-mass halo is simply not large enough to provide a significant cross section. The contribution from haloes with $M_\mathrm{halo}>10^{11}$~M$_\odot$ is dominant and steeply increasing at $N_\mathrm{H\,\textsc{i}}>10^{21}$~cm$^{-2}$. The fact that low- (high-) mass haloes dominate low (high) column density DLAs agrees qualitatively with \citet{Pontzen2008}, although they found that haloes more massive than $10^{10.5}$~M$_\odot$ only dominate for $N_\mathrm{H\,\textsc{i}}>10^{21.5}$~cm$^{-2}$. Unresolved haloes ($M_\mathrm{halo}<10^9$~M$_\odot$) cannot dominate the absorption for $N_\mathrm{H\,\textsc{i}}>10^{17}$~cm$^{-2}$, because most of it is already accounted for by resolved haloes.

We also split the gas into several SFR bins, but do not show the result. We found that for $N_\mathrm{H\,\textsc{i}}<10^{21}$~cm$^{-2}$, the largest contribution (about 30 per cent), comes from haloes with $0.01< {\rm SFR} <0.1$~M$_\odot /{\rm yr}$. For higher column densities  objects with ${\rm SFR} >10$~M$_\odot /{\rm yr}$ are most important, accounting for 30--65 per cent of the H\,\textsc{i} absorption, depending on the H\,\textsc{i} column density.

At all column densities, most H\,\textsc{i} gas is inflowing, but there are important and roughly equal contributions from static and outflowing gas (bottom left panel). The contributions from inflowing, static, and outflowing gas are nearly constant (about 55, 30, and 15 per cent, respectively) up to $N_\mathrm{H\,\textsc{i}}=10^{21}$~cm$^{-2}$, at which point the contribution of static gas increases relative to those of both in- and outflows. If we separate the gas into inflowing or outflowing without any minimum radial velocity threshold (not shown), then the ratio of inflowing to outflowing remains approximately the same, with 65 per cent (35 per cent) of the H\,\textsc{i} absorption due to inflowing (outflowing) gas.

Above $N_\mathrm{H\,\textsc{i}}=10^{18}$~cm$^{-2}$, the contribution to H\,\textsc{i} absorption from gas that is \textit{in haloes} and inflowing, static, or outflowing (bottom middle panel) is almost equally large as that from \textit{all} gas that is inflowing, static, or outflowing (bottom left panel). Below $N_\mathrm{H\,\textsc{i}}=10^{18}$~cm$^{-2}$, their contribution is lower, because many absorbers reside outside of haloes.

The fraction of H\,\textsc{i} absorption due to gas inside galaxies, i.e.\ the ISM, is only significant for DLAs and dominates at $N_\mathrm{H\,\textsc{i}}\gtrsim10^{21}$~cm$^{-2}$ (bottom right panel). Two-thirds of the $10^{17.5}<N_\mathrm{H\,\textsc{i}}<10^{21}$~cm$^{-2}$ material is not participating in star formation yet, but will do so within the next 1.2~Gyr. About half of this gas will accrete onto a galaxy for the first time. Gas previously ejected from galaxies contributes a comparable amount of absorption as gas accreting onto galaxies for the first time.

\section{Discussion and conclusions} \label{sec:concl}

After post-processing with radiative transfer, the OWLS reference simulation matches the observed $z=3$ H\,\textsc{i} column density distribution over ten orders of magnitude in $N_\mathrm{H\,\textsc{i}}$ \citep{Altay2011}. This success gives us confidence that we can use this simulation to study the relation between cold accretion and high column density H\,\textsc{i} absorbers.

Like other simulations, our simulation also shows cold streams and filaments inside and outside of haloes \citep{Voort2011a}. Gas in many of these streams does not go through an accretion shock near the virial radius and, because it is able to accrete onto a galaxy much more efficiently than the hot, diffuse gas, it is vital for fuelling star formation \citep[e.g.][]{Keres2005, Ocvirk2008, Keres2009a, Voort2011a, Voort2011b, VoortSchaye2011, Faucher2011}. Here, we demonstrated that our simulation would not have matched the observed H\,\textsc{i} column density distribution without cold accretion. This fact alone provides evidence that cold accretion has already been observed. 

Motivated by the theoretical framework of cold, filamentary accretion, we have investigated the importance of gas that satisfies certain selection criteria for column densities above $N_\mathrm{H\,\textsc{i}}=10^{16}$~cm$^{-2}$ at $z=3$. It is difficult to check the morphology of all the H\,\textsc{i} in the simulation, but from images (see also Figure \ref{fig:image}) we do know that a lot of the accreting cold-mode gas is filamentary. However, the highest column density gas tends to be clumpy. Even though much of this gas is infalling and has never gone through a virial shock, it would not all be classified as streams based on its morphology.  

We will first discuss the results separately for three column density regimes: i) the Lyman-$\alpha$ forest; ii) LLSs and low column density DLAs, and iii) high column density DLAs. We will then discuss our results in the context of the cold-mode accretion paradigm, compare them to recent theoretical results from \citet{Fumagalli2011}, and conclude.

\begin{enumerate}

\item Nearly all H\,\textsc{i} absorbing gas in the denser part of the Lyman-$\alpha$ forest ($N_\mathrm{H\,\textsc{i}}=10^{16-17}$~cm$^{-2}$) has never gone through a virial shock ($T_\mathrm{max}<10^{5.5}$~K) and most of it is outside of haloes (the same is true for the low column density forest, but we did not show this here). Even though 60 per cent is flowing towards the nearest halo (nearest in units of $R_\mathrm{vir}$) with velocities greater than a quarter of its circular velocity, only 15--40 per cent will participate in star formation before $z=2$, i.e.\ within 1.2~Gyr. Most of the Lyman-$\alpha$ forest gas that will accrete onto a galaxy before $z=2$ will do so for the first time. 

\item Nearly all H\,\textsc{i} absorbing gas in LLSs ($10^{17}<N_\mathrm{H\,\textsc{i}}<10^{20.3}$~cm$^{-2}$) and in low column density DLAs ($10^{20.3}<N_\mathrm{H\,\textsc{i}}<10^{21}$~cm$^{-2}$) has never gone through a virial shock. About 60--95 per cent of the H\,\textsc{i} in LLSs  is inside haloes with larger contributions from lower-mass haloes: 30--40 per cent from $10^{9-10}$~M$_\odot$, 20 per cent from $10^{10-11}$~M$_\odot$, 10--25 per cent from $10^{11-12}$~M$_\odot$, and less than 15 per cent from $10^{12-13}$~M$_\odot$ haloes. Even inside haloes, almost half of the neutral gas is inflowing with velocities greater than 25 per cent of the circular velocity. For LLSs, more than 90 per cent of the H\,\textsc{i} is currently outside of galaxies, but 40--70 per cent will accrete onto a galaxy within 1.2~Gyr and just over 30 per cent will do so for the first time. The gas reservoir that is feeding galaxies is thus, at least in part, associated with LLSs. A significant fraction of the gas in LLSs and low column density DLAs has previously already been inside a galaxy. This agrees with the finding that the re-accretion of gas that has been ejected is important for the build-up of galaxies \citep{Oppenheimer2010}.

\item About 10--30 per cent of H\,\textsc{i} absorbing gas in high column density DLAs ($N_\mathrm{H\,\textsc{i}}>10^{21}$~cm$^{-2}$) was heated above $10^{5.5}$~K in the past. This gas has either been accreted in the hot mode or it has been shock-heated by a galactic wind. Depending on the column density, 55--95 per cent of the high-$N_{\rm H\,\textsc{i}}$ DLA gas is contained in the ISM of galaxies and 50--70 per cent of the H\,\textsc{i} is inside haloes more massive than $10^{11}$~M$_\odot$. Outflowing gas contributes about 15 per cent of the H\,\textsc{i}. Inflowing gas provides the largest contribution to the H\,\textsc{i} of high column density DLAs, but the static gas component approaches the importance of the inflowing gas at the highest column densities. 

\end{enumerate}

The simplest way of defining cold-mode accretion is purely based on its maximum past temperature. Gas with temperatures below $10^{5.5}$~K is able to cool efficiently, because the cooling function peaks at $T\approx10^{5-5.5}$~K \citep[e.g.][]{Wiersma2009a}. Above this temperature, cooling times become longer. We define cold-mode gas as having $T_\mathrm{max}<10^{5.5}$~K. Using this definition, practically all H\,\textsc{i} absorption is due to cold-mode gas. This is not so surprising for the Lyman-$\alpha$ forest, which is comprised mostly of gas outside of haloes and would therefore not be expected to have been shock-heated to high temperatures. LLSs are dominated by low-mass haloes with $M_\mathrm{halo}<10^{11}$~M$_\odot$ and $T_\mathrm{vir}<10^{5.4}$~K. Even if this gas were heated to the virial temperature, we would still classify it as cold-mode accretion. However, most neutral gas in haloes more massive than $10^{11}$~M$_\odot$ cannot have been heated to the virial temperature, or we would have found that the fraction of the H\,\textsc{i} with $T_\mathrm{max}>10^{5.5}$~K exceeds the fraction of the H\,\textsc{i} contributed by haloes with mass $>10^{11}$~M$_\odot$, which increases from about 20 to more than 95 per cent from $N_\mathrm{H\,\textsc{i}}=10^{18}$ to $10^{22}$~cm$^{-2}$. However, we find that the actual fraction of H\,\textsc{i} contributed by gas with $T_\mathrm{max}>10^{5.5}$~K is only 5--30 per cent for this column density range. Thus, it is clear that even for high column density absorbers, the H\,\textsc{i} content is dominated by gas that has never gone through an accretion shock near the virial radius.

We can also think of cold-mode accretion as gas that has already accreted onto haloes, but not yet onto galaxies, without having reached temperatures above $10^{5.5}$~K, flowing in with relatively high velocities, and that will accrete onto a galaxy in the near future. About 30 per cent of all $10^{18}<N_\mathrm{H\,\textsc{i}}<10^{20.5}$~cm$^{-2}$ systems satisfy all these criteria. Therefore, even this set of highly restrictive criteria leads to the conclusion that cold-mode accretion has already been observed in the form of high column density H\,\textsc{i} absorption systems.

However, if we think of cold-mode accretion as accreting onto galaxies for the first time, in addition to having low $T_\mathrm{max}$, being inside haloes, and rapidly inflowing, the fraction of LLSs satisfying this requirement is about a factor of two lower, but still around 15 per cent. Cold streams also contain galaxies, so the gas that is being recycled could still be accreting along these streams. If we include previously ejected gas in our definition of cold-mode accretion, the metallicity of cold-mode gas would not necessarily be low.

\citet{Fumagalli2011} already pointed out that cold streams have likely been detected as LLSs. They used high-resolution simulations of seven galaxies with $10^{10.7}<M_\mathrm{halo}<10^{11.5}$~M$_\odot$ at $z\approx3$ by resimulating a cosmological simulation. They included ionizing radiation from local stellar sources, which we did not. \citet{Nagamine2010} have shown that local stellar sources change the column density distribution function by less than 0.1~dex, although \citet{Fumagalli2011} find a somewhat larger effect (0--0.5~dex). \citet{Fumagalli2011} investigated H\,\textsc{i} absorption from ``central galaxies'', which they defined as all gas within $0.2R_\mathrm{vir}$, and from ``streams'', defined as all gas at radii $0.2R_\mathrm{vir}<R<2R_\mathrm{vir}$, and found that ``cold streams'' mostly have $10^{17}<N_\mathrm{H\,\textsc{i}}<10^{19}$~cm$^{-2}$. However, their definition of ``cold streams'' was all gas at $0.2R_\mathrm{vir}<R<2R_\mathrm{vir}$, which excludes neither gas that has been shock-heated to the virial temperature, i.e. hot accretion, nor gas that is outflowing.

The halo sample of \citet{Fumagalli2011} underproduces $N_\mathrm{H\,\textsc{i}}<10^{17}$~cm$^{-2}$ systems, which they argue is because many of these absorbers live outside of haloes. This is in agreement with our finding that the contribution of halo gas is strongly declining with decreasing $N_\mathrm{H\,\textsc{i}}$, although we find that up to 30 per cent still stems from low-mass haloes not covered by their simulations. We find that for $10^{16}<N_\mathrm{H\,\textsc{i}}<10^{20}$~cm$^{-2}$, haloes with $10^9<M_\mathrm{halo}<10^{10}$~M$_\odot$ account for about as much H\,\textsc{i} absorption as all higher-mass haloes combined. The fact that $10^{18}<N_\mathrm{H\,\textsc{i}}<10^{20}$~cm$^{-2}$ systems are also underproduced in their sample is therefore consistent with our finding that very low-mass haloes dominate the H\,\textsc{i} column density distribution function. \citet{Fumagalli2011} conclude that haloes in the mass range $10^{10}<M_\mathrm{halo}<10^{12}$~M$_\odot$ already account for 20--30 per cent of the LLSs and DLAs. We have found this fraction to be even larger: 30--40 per cent for LLSs and 40--70 per cent for DLAs. This difference could be due to their inclusion of ionizing radiation from local sources.

Our results lead to the somewhat stronger conclusion that for $10^{18}<N_\mathrm{H\,\textsc{i}}<10^{21}$~cm$^{-2}$ more than 80 per cent of the H\,\textsc{i} absorption is caused by cold-mode halo gas. Only about 5 per cent is caused by hot-mode gas. The remainder may arise from the high-density part of the IGM or from haloes with $M_\mathrm{halo}<10^9$~M$_\odot$. At lower column densities ($N_\mathrm{H\,\textsc{i}}<10^{18}$~cm$^{-2}$), the contribution of halo gas declines, though more than 95 per cent of the absorbing gas is still cold mode. For higher column densities ($N_\mathrm{H\,\textsc{i}}>10^{21}$~cm$^{-2}$), the absorption is dominated by the ISM of galaxies in haloes with $M_\mathrm{halo}>10^{11}$~M$_\odot$ and, depending on the column density, the mean cold-mode contribution is between 70 and 90 per cent. At all column densities, inflowing gas accounts for most of the H\,\textsc{i} column density, but there are significant contributions from static and outflowing gas.

We conclude that cold streams are real. They have already been observed in the form of high column density absorbers, mainly in systems with $10^{17}<N_\mathrm{H\,\textsc{i}}<10^{21}$~cm$^{-2}$. Cold flows are critical for the success of our simulation in reproducing the observed $z=3$ column density distribution of damped Lyman-$\alpha$ and particularly that of Lyman limit systems.

\section*{Acknowledgements}

We would like to thank Olivera Rakic and all the members of the OWLS team for valuable discussions and Claudio Dalla Vecchia and Alireza Rahmati for helpful comments on an earlier version of the manuscript. The simulations presented here were run on the Cosmology Machine at the Institute for Computational Cosmology in Durham as part of the Virgo Consortium research programme. The ICC Cosmology Machine is part of the DiRAC Facility jointly funded by STFC, the Large Facilities Capital Fund of BIS, and Durham University. This work was sponsored by the National Computing Facilities Foundation (NCF) for the use of supercomputer facilities, with financial support from the Netherlands Organization for Scientific Research (NWO), also through a VIDI grant, and from the Marie Curie Initial Training Network CosmoComp (PITN-GA-2009-238356).

\bibliographystyle{mn2e}
\bibliography{HIabsorption}

\bsp

\label{lastpage}

\end{document}